\documentclass[%
 onecolumn,
 amsmath,amssymb,
 aps,
]{revtex4-2}

\usepackage{graphicx}
\usepackage{dcolumn}
\usepackage{bm}
\usepackage{amsmath}

\newcommand{\uu}{\mathbf{u}}
\newcommand{\ub}{\bar{\mathbf{u}}}

\newcommand{\xp}{\mathbf{x}_p}
\newcommand{\up}{\mathbf{U}_p}

\newcommand{\p}{\partial}
\begin{document}

\preprint{APS/123-QED}

\title{Inertial Migration In Micro-Centrifuge Devices}
\author{Samuel Christensen}
\email{sam.em.chris@gmail.com}
\author{Marcus Roper}%
 
\affiliation{%
 UCLA Department of Computational Medicine\\
 UCLA Department of Mathematics
}%


\date{\today}

\begin{abstract}
Within microcentrifuge devices, a microfluidic vortex separates larger particles from a heterogeneous suspension using inertial migration, a phenomenon that causes particles to migrate across streamlines. The ability to selectively capture particles based on size differences of a few microns makes microcentrifuges useful diagnostic tools for trapping rare cells within blood samples. However, rational design of microcentrifuges has been held back from its full potential by a lack of quantitative modeling of particle capture mechanics. Here we use an asymptotic method, in which particles are accurately modeled as singularities in a linearized flow field, to rapidly calculate particle trajectories within microcentrifuges. Our predictions for trapping thresholds and trajectories agree well with published experimental data. Our results clarify how capture reflects a balance between advection of particles within a background flow and their inertial focusing and shows why the close proximity of trapped and untrapped incoming streamlines makes it challenging to design microcentrifuges with sharp trapping thresholds.

\end{abstract}

\maketitle


\medmuskip=0mu
\thinmuskip=0mu
\thickmuskip=0mu
\section{Introduction}
Microcentrifuges are a recently developed class of microfluidic devices that can be used to selectively trap large particles from flowing suspensions. The devices consist of a series of chambers connected with microfluidic channels. Within each chamber, one or more eddies may form. \emph{Inertial migration} causes particles in moderate Reynolds number flows to travel across streamlines. Larger particles migrate faster and are more likely to become trapped within the microcentrifuge chamber \cite{khojah2017size, haddadi2017inertial}. Size-based trapping may be used to trap the largest particles within the suspension, and shows promise as a tool for analyzing cell types in a patient blood sample, e.g. for isolating large circulating cancer cells from small red blood cells \cite{che2016classification,dhar2018evaluation,chen2012microfluidic}.
\begin{figure}
    \centering
    \includegraphics[width=0.75\columnwidth]{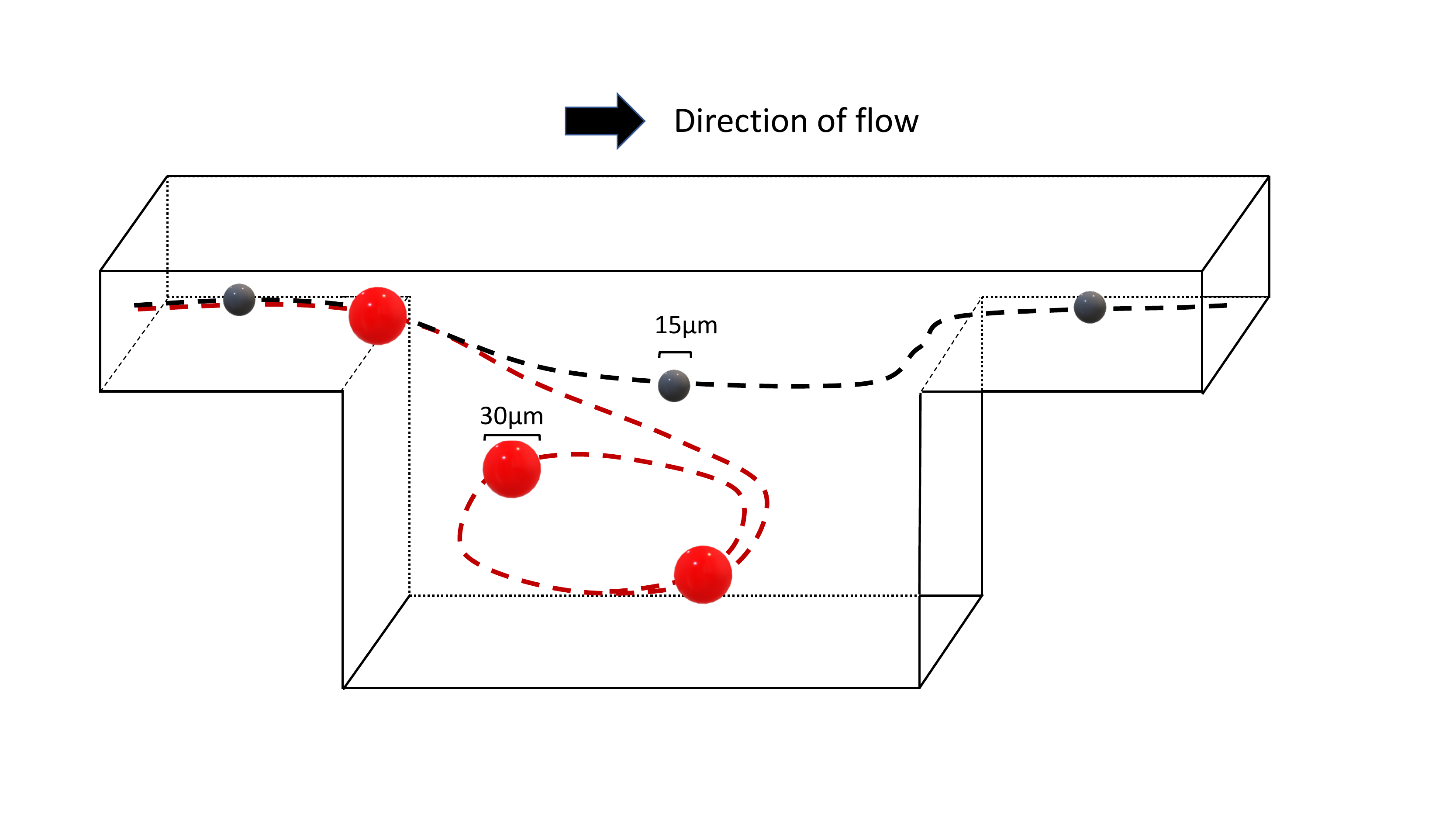}
    \caption{A diagram of the microcentrifuge (not to scale). Microfluidic channels lead to the microcentrifuge chamber. At sufficient Reynolds numbers, a fluid eddy forms in the chamber and inertial migration pushes larger particles into the chamber, where they are then trapped in the eddy, while smaller particles are not captured and continue through the device.}
    \label{fig:diagram}
\end{figure}


The range of possible particle behavior within the complex three dimensional flows that occur in microcentifuge chamber is not well understood, nor is there, to our knowledge, has any straightforward mechanistic description of how a microcentrifuge operates. In addition to the intrinsic interest of identifying new inertial microfluidic phenomena, mechanistic explanation of microcentrifuge function opens a door to solving the reverse problem, of designing a chamber or channel geometry to target a particular size threshold for trapping.

A large impediment to this understanding is how much less well developed the theory of inertial microfluidic migration is, relative to theories modeling the behaviors of particles in zero Reynolds number flows. Although inertial focusing starts at arbitrarily low Reynolds numbers, in practical implementations for useful trapping thresholds to be accessed, the devices so far built have all operated at Reynolds numbers between 50 and 300. Inertial migration is caused by the rigid particle disrupting the otherwise smoothly varying fluid flow throughout the channel. Although the physics of inertial migration in (uni-directional) pipe flow has been studied extensively, (see e.g. \cite{schonberg1989inertial} and \cite{hood_lee_roper_2015}): the key difference between pipe flow and more complex flow is the varying flow that the particle experiences as it advects through the channel. In this paper, we develop an asymptotic theory that reduces the calculation of particle migration velocity to solving a quasi-steady, linear problem.

Direct numerical simulation of particle trajectories within inertial microfluidic devices requires solving for the motion of particles suspended in a fluid-filled domain whose boundaries constantly change due to the movement of the particles. Since, in practical examples, channels operate at moderate Reynolds number, between 5-200, flows cannot be modeled by numerical methods that are designed for small Reynolds number particle-flows such as Stokesian dynamics \cite{sierou2001accelerated} or boundary integral methods \cite{liron1992motion}. The time evolving geometry of migrating particles and nonlinear terms favors numerical methods such as immersed boundary \cite{kazerooni2017inertial,nakagawa2015inertial} or immersed interface \cite{xu2020octree} which embed moving boundaries within a fixed computational grid. Both methods afford a lot of freedom in the choice of numerical method for solving the Navier-Stokes equations; in inertial microfluidic simulations the Lattice-Boltzmann method (LBM) is a popular method \cite{bazaz2020computational} and has been used for calculating particle migration \cite{haddadi2017inertial,sun2016three,chun2006inertial}. The Force Coupling Method\cite{loisel2013effect} replaces particles with forcing terms which enforce rigid body motion on a finite volume of fluid to emulate particle dynamics and has been used to calculate inertial migration\cite{abbas2014migration}. Although existing numerical simulations have illuminated the physics of inertial focusing, the high computational cost of nonlinear 3D simulations has meant that predictive simulations are not currently used to design or optimize inertial microfluidic devices.


In our simplification of the physics of inertial migration, a 3D linear PDE models moderate Reynolds number fluid flow and supplies a vector field representing inertial migration that may be added to the background flow. This simplification allows for easier understanding of the various sorting techniques that are used by microfluidic devices. In this study, we will first develop our equations for particle motion and then we will compare our theoretical results with data from a microfluidic device separating different cell types suspended in blood along with differently sized plastic beads.

\section{Mathematical Methods}
Inertial migration is caused by the disturbance the particle causes to the background flow, the disturbance flow $\uu' = \uu - \ub$ is the difference between the flow with the particle, $\uu$, and the flow without the particle,  (called the background flow here) $\ub$. We get our main equations by plugging these definitions into the Navier-Stokes equations and non-dimensionalizing by the speed scale $U$ and length scale $L$ of the background flow:
\begin{eqnarray}
    \Delta \uu' - \nabla p' &=& Re\left( \frac{\partial \uu '}{\partial t} + \ub \cdot \nabla \uu ' + \uu'\cdot \nabla \ub  + \uu' \cdot \nabla \uu' \right )  \label{eq:NSE} \\
    \nabla \cdot \uu' &=& \mathbf{0} \nonumber \\
    \uu' &=& \up-\ub(\xp) +\Omega_p \times (\mathbf{x} - \xp ) \qquad\text{on }|\mathbf{x}-\xp|=a/L \label{eq:particleBC}
\end{eqnarray}
In addition to $\uu' = 0$ on the walls of the channel. Neither boundary conditions nor the incompressibility equation are altered by following transformations of our equation, so we focus on the different incarnations of the momentum balance equation, Eq. \ref{eq:NSE}. Here $Re=\frac{U L \rho}{\mu}$ is the channel Reynolds number, and is typically in the range 50-300, both in real experiments and our calculations.
The disturbance velocity is generated by the boundary condition on the sphere. The particle can translate and rotate with the fluid, but it resists the shearing motion of the fluid near its surface, setting up the disturbance velocity $\uu'$. Near the particle, the velocity profile is approximately that of simple shear; $\ub \approx \ub(\xp) + \boldsymbol{\gamma} \cdot (\mathbf{x}-\xp)$, where $\gamma_{ij} \equiv \frac{\partial \bar{u}_i}{\partial x_j}$, and we may further decompose the linearized flow field into rotational and straining components: $\ub \approx \ub(\xp) + \mathbf{\omega} \times (\mathbf{x}-\xp) + \mathbf{E}\cdot (\mathbf{x}-\xp)$, where $\omega_i = -\frac{1}{2}\epsilon_{ijk}\gamma_{jk}$, with $\mathbf{\epsilon}$ the unit alternating tensor, and $E_{ij} = \frac{1}{2}(\gamma_{ij} + \gamma_{ji})$. Setting $\up \approx \ub(\xp)$ and $\Omega_p = \omega$ renders the particle force and torque free at first order.

We now know the size of all the components involved in right hand side of Eq. \ref{eq:NSE}: under our non-dimensionalization, both $\uu'$ and $\ub$ are of size $O(\frac{a}{L})$ near the particle. The size of $\frac{\p \uu'}{\p t}$ is related to the boundary condition Eq. \ref{eq:particleBC} who's size is the same as $\frac{\p \mathbf{E}(\xp(t))}{\p t } = \up \cdot \nabla \mathbf{E}$ which is small as long as the length scale at which $\mathbf{E}$ changes is large compared to $a$. Therefore the right hand side is size $\frac{a^2}{L^2} Re$. The size ratio $\alpha := \frac{a}{L}\ll 1$ is small enough that we assume $\alpha^2 Re<1$, which forms the core of our asymptotic expansion. Near the particle, we assume shear dominates and set the inertial terms to zero. Suppressing these inertial terms, we arrive at: $\Delta \uu'-\nabla p' = 0$, along with the usual incompressibility and boundary conditions. This problem of Stokes' flow around a sphere has solution:
\begin{equation}
    \uu' = -\mathbf{E}:\frac{5(\mathbf{x}-\xp)(\mathbf{x}-\xp)(\mathbf{x}-\xp)}{|\mathbf{x}-\xp|^5} +O(|\mathbf{x}-\xp|^{-4}) \label{eq:stresslet}
\end{equation}
\cite{kim2013microhydrodynamics}. However, this solution is not consistent with our complete neglect of the inertial terms from Eq. \ref{eq:NSE}, because as $|\mathbf{x}-\xp|\to \infty$, $\Delta \uu' \sim 1/r^4$, while $Re\, \ub\cdot \nabla \uu'\sim Re/r^2$, becomes co-dominant with viscous stresses when $|\mathbf{x}-\xp|\sim Re^{-1/2}$. We therefore posit that the flow contains an outer region in which inertial terms may not be neglected\cite{schonberg1989inertial}. However, since decay of the disturbance velocity means that $|\uu'|\ll |\ub|$ within this region, we may linearize the inertial terms within this outer region.

In the outer region, we model the particle as a moving singularity. Mathematically, this means replacing the rigid body motion of Eq. \ref{eq:particleBC} with a forcing term equal to $F(\mathbf{x}) =-\frac{20\pi}{3} E(\xp):\nabla \delta(\mathbf{x}-\xp(t))$. Instead of working with a moving particle, we approximate this by looking for a traveling wave solution of the form $\uu'(\mathbf{x}-\xp(t))$, this approximation transforms the time derivative into $\frac{\p \uu'(\mathbf{x}-\xp(t))}{\p t} = \up \cdot \nabla \uu$ which we will approximate as $\frac{\p \uu'(\mathbf{x}-\xp(t))}{\p t} = \ub(\xp) \cdot \nabla \uu$.


Our solution in the inner region does not predict particle migration at leading order; instead the stresslet disturbance field modeled in Eq. \ref{eq:stresslet}, forces the outer region disturbance velocity. This forcing can be represented equivalently as a boundary condition as $\mathbf{x}\to \xp$, or, directly, by introducing a force dipole term within the equation:

 \begin{eqnarray}
      \nabla \uu' -\nabla p' &=& Re\left((\ub - \ub(\xp)) \cdot \nabla \uu' + \uu' \cdot \nabla \ub \right) - \frac{20\pi}{3}E:\nabla \delta(\mathbf{x}-\xp) \label{eq:Oseen}\\
      \nabla\cdot \uu' &=& 0 \nonumber
 \end{eqnarray}
 
The migration velocity is found by solving this PDE and evaluating it at the location of the particle $\mathcal{M}(\xp) = \uu'_{\xp}(\xp)$, where the subscript denotes that the singularity was located at $\xp$ according to Eq. \ref{eq:Oseen}. We then advect the particle according to following differential equation
\begin{eqnarray}
    \frac{d\xp}{dt} = \ub (\xp) + \alpha^3\mathcal{M}(\xp) +\frac{\alpha^2}{6}\Delta \ub (\xp) \label{eq:ode}
\end{eqnarray}
This equation includes the term for Faxen's Law, which is the finite size particle correction for force free advection of particles in $Re=0$ flows. This formula is exact for unbounded $Re=0$ flows, but in our use is accurate to the order $O(Re^0 \alpha^4)$.

When calculating the trajectories of particles, we pre-calculated the migration velocity at 450 points throughout the three dimensional channel by Eq. \ref{eq:ode}. We then used linear interpolation to extend our particle velocity calculation through the entire domain.

Eq. \ref{eq:ode} can be used to to advect particles throughout the channel, however its accuracy is limited to conditions where the underlying asymptotic expansion is valid. In straight channels it has been shown that this asymptotic approximation is accurate for $\alpha^2 Re = O(1)$\cite{schonberg1989inertial}, however both the size and background flow speed of the microfluidic device can change significantly depending on if we are in a tight channel or large chamber, eg. the large notched chamber in the microcentrifuge.  if we define the Reynolds number using the average flow rate in a cross section with length $L$, we see that $Re = \frac{\int_A\ub dS L}{L^2\nu}=\frac{C}{\nu L}$ where C is the flow rate of the device and is constant for all cross sections in the channel. In this calculation we see that the Reynolds number decreases like $\frac{1}{L}$ in larger channels, so the asymptotic expansion that forms Eq. \ref{eq:ode} is more accurate in the larger testing chambers than it is in the smaller channels leading up to it.

Another potential source of error in the asymptotic approximation is our assumption of the traveling wave solution. While $\frac{\p \uu'(\mathbf{x}-\xp(t))}{\p t} = \ub(\xp) \cdot \nabla \uu$ is exact in the bulk flow, the particle is moving with respect to the walls of the device, which will cause additional changes to the flow as $\xp$ moves. It can be shown that this error is proportional to $\up \cdot \nabla \uu'|_{\text{walls}}$ which is small provided the particle is either far away from the wall or not headed directly at it.

\section{Analysis of Microcentrifuge}
\subsection{Background Flow Patterns at Different Reynolds numbers}
In \cite{khojah2017size}, Khojah et al. observed a microcentrifuge trapping differently sized cells from blood samples. They found that at different Reynolds numbers it preferentially captured different sizes of cells. At $Re=125$ the micro centrifuge consistently captured larger particles and smaller particles could pass through. At $Re=175$ they found that the microcentrifuge would inconsistently capture particles of all sizes. At $Re=225$ the microcentrifuge would capture smaller particles with more consistency than it would capture larger particles.

\begin{figure}
    \hspace*{-1.3cm}
    \includegraphics[width=1\columnwidth]{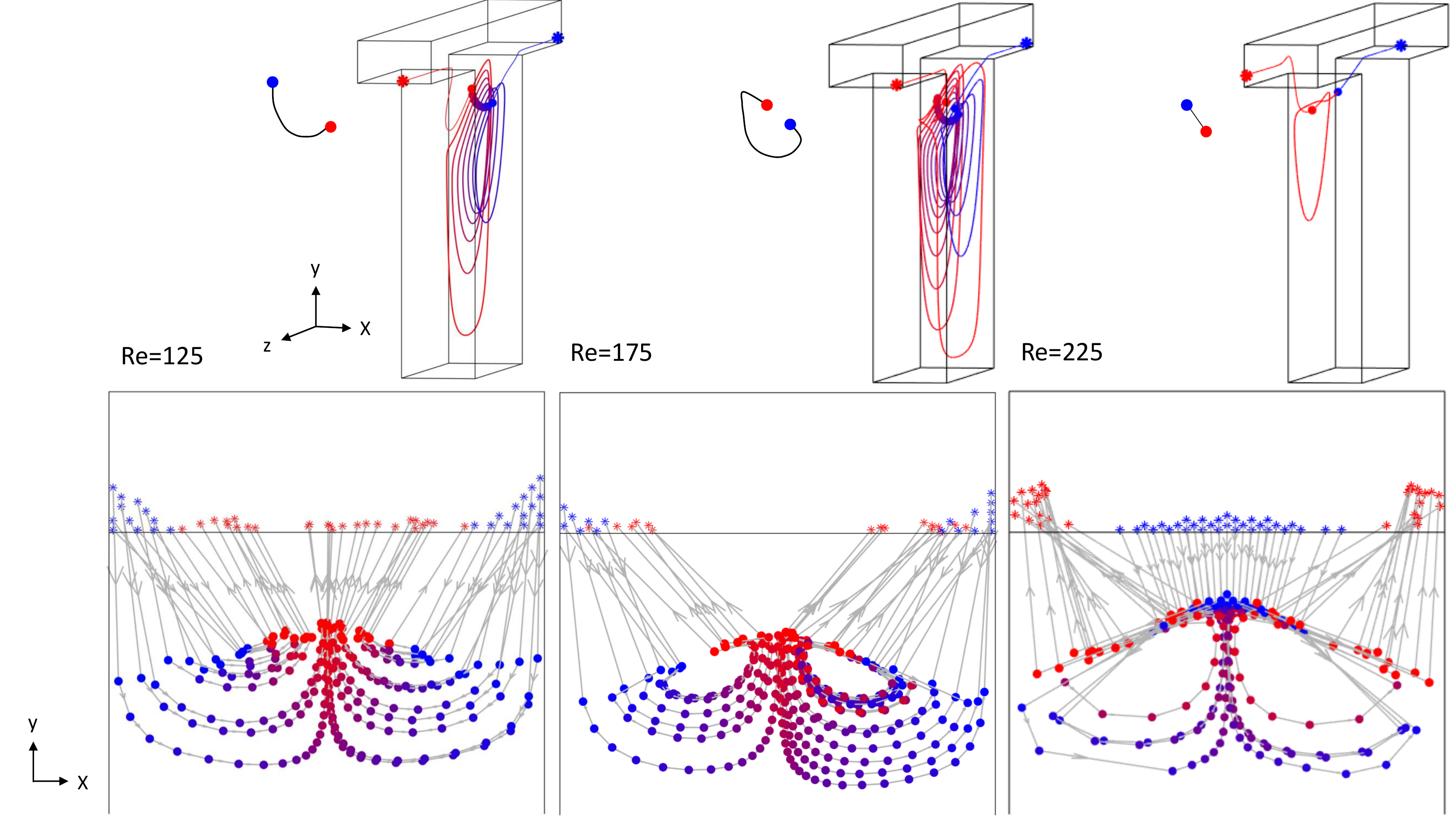}
    \caption{The fluid flow within the microcentrifuge experiences a topological change as Re changes from 125 to 225. The top row shows an example streamline and the bottom row shows the Poincar\'e section of many streamlines and how they progress through the channel. Blue shows where the streamlines enter and red shows where the streamlines exit. The trajectory cartoon demonstrates how the entrance and exit of the streamlines loop back on themselves and change their locations in the channel, representing a topological change of how the fluid moves through the channel.}
    \label{fig:BGflow}
\end{figure}

 In numerical studies using COMSOL Multiphysics \cite{multiphysics1998introduction}, we found that these Reynolds numbers correspond with major changes to the background flow. Streamlines were calculated for Reynolds numbers 125, 175, and 225 were calculated using P2+P1 tetrahedral elements at the 'extremely fine' mesh size setting.

Fig. \ref{fig:BGflow} shows how the way the fluid flows through the microcentrifuge chamber changes as we increase the Reynolds number. The top row shows 3D visualization of example particle trajectories and the bottom row shows via Poincar\'e section how the streamlines travel through the chamber once they are caught, the dots represent where the trajectories intersect a plane at the top of each of the orbits the streamlines performed within the chamber. Blue represents where the streamlines entered and red represents where they exited the chamber. The x- and y- coordinates of streamlines entering and exiting the chamber are shown with asterisks instead of dots.

In Fig. \ref{fig:BGflow}, we see that at $Re=125$ the fluid flows through the microcentrifuge chamber by having the corner streamlines flow down and into the chamber, then the fluid streamlines rotate towards the center of the chamber and then spiral outwards, eventually exiting the chamber. At $Re = 175$ the path of the streamlines is looping back on itself and is no longer consistently focusing towards the center of the channel. At $Re=225$, the loop has inverted: instead of flowing from out to in, the streamlines enter the chamber from the center of the channel and leave out the sides. Inset diagram in the top row of Fig 2. show the exchange of starting and ending streamline locations as $Re$ is increased.

\subsection{Particle Trajectories}

\begin{figure}
    \centering
    \includegraphics[width=\columnwidth]{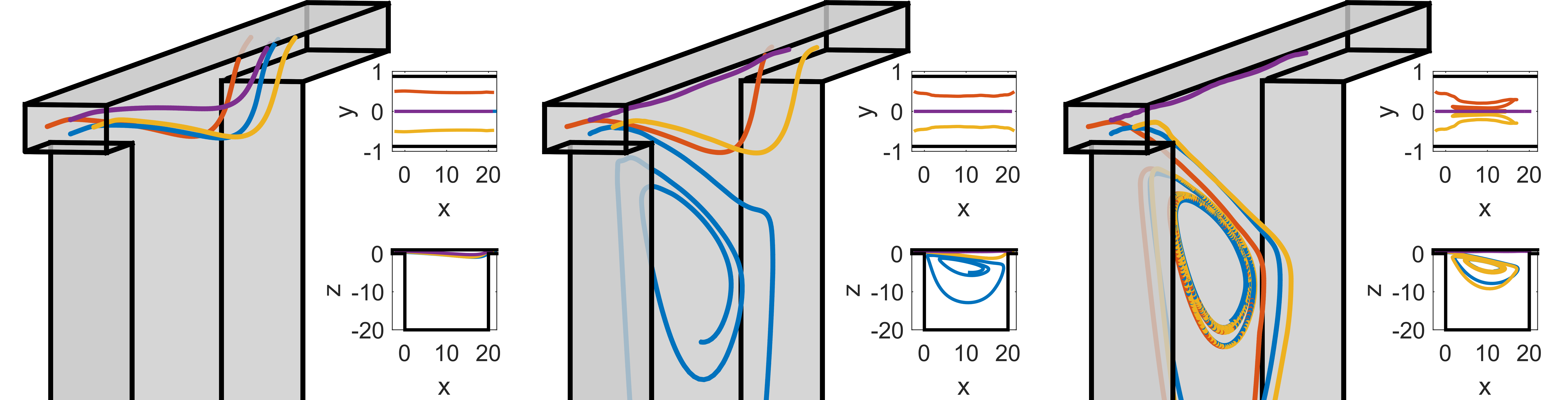}
    \caption{The trajectories of background flow (left), $24\mu m$ diameter particles (middle), and $28\mu m$ diameter particles. Small particles follow the background flow and do not enter the microcentrifuge chamber, capture is predicted to begin at approximately $22.9 \mu m$, but only particles along the bottom focusing position are captured. For larger particles, particles along all but the top focusing position are captured. All captured particles stably focus towards the mid plane at this Reynolds number.}
    \label{fig:3Dfig}
\end{figure}
We will now start looking at the microcentrifuge capture particles, but first we need to look into the initial condition. Before the particle reaches the microcentrifuge, it will travel through a narrow rectangular channel (in our case $40\mu $m$\times 70\mu$m), at $Re=125$ the particles gather along 4 stable focusing streamlines near the center of each channel wall \cite{hood_lee_roper_2015, di2007continuous}. This means the particles will be tightly grouped along 4 focusing streamlines before they enter the channel, with the majority of particles evenly split between the two larger walls of the channel (corresponding to the blue and purple trajectories in Fig. \ref{fig:3Dfig} \cite{christensen2022fast}). 

In Fig \ref{fig:3Dfig}, we place differently sized particles along the 4 focusing streamlines and advance their position using Eq. \ref{eq:ode}. All 4 focusing streamlines do not enter the notch, at this Reynolds number the streamlines that enter the channel are along the bottom corners \cite{https://doi.org/10.48550/arxiv.2101.07242}, however at particle size $24 \mu$m  particles along the bottom streamline migrate away from the streamline enough that they become captured in the eddy within the notch. at particle size $28\mu $m the particles along the minor streamlines are also captured. At particle size $35\mu $m the particles along the top streamline are captured as well, however these particles are $87.5\%$ of the channel height and the identification of 2 distinct focusing streamlines may no longer be relevant. Our calculations also show that above above diameter $28\mu $m, the particles from the minor streamlines will quickly focus towards the mid plane \ref{fig:3Dfig}. Capture of particles along the minor axis is driven by the strong inertial migration towards the microvortex, inertial focusing actually fights the particle from migrating towards the center of the channel but is overpowered by the background flow and the particle is ultimately driven towards the mid plane,  The rapid convergence of all particles toward the symmetry plane agrees with numerical results from \cite{haddadi2017inertial}.

Changes in the background fluid flow recorded in section A completely change the way particle capture is achieved and explain the differences in what size particles are captured by the micro-centrifuge. At $Re=125$, particles are captured when inertial migration pushes the particle into the microcentrifuge chamber, the limit cycle is created as a product of the balance between the fluid flow spiraling the particle upwards and inertial migration pushing the particle downwards at the top of the spiral. At $Re=175$ all particles are predicted to enter the microcentrifuge chamber, but it is not well understood what keeps them in the close to turbulent eddy that is spiraling within the chamber. We can integrate the physics of capture with the description in section A of the three dimensional paths of streamlines through the chamber.

Particles that enter the microcentrifuge chamber continue to experience inertial migration within that chamber. At $Re=125$, the streamlines that particles follow, absent inertial migration, spiral outwards within the mid plane Fig. \ref{fig:BGflow}, while inertial migration points downward into the chamber. The balance of these two effects caused particle trajectories to converge to a limiting orbit, shown in Fig. \ref{fig:3Dfig}. We report in more detail upon the limiting orbit in section C.

\subsection{Comparison With Data}
\begin{figure}
    \centering
    \includegraphics[width=1.0\columnwidth]{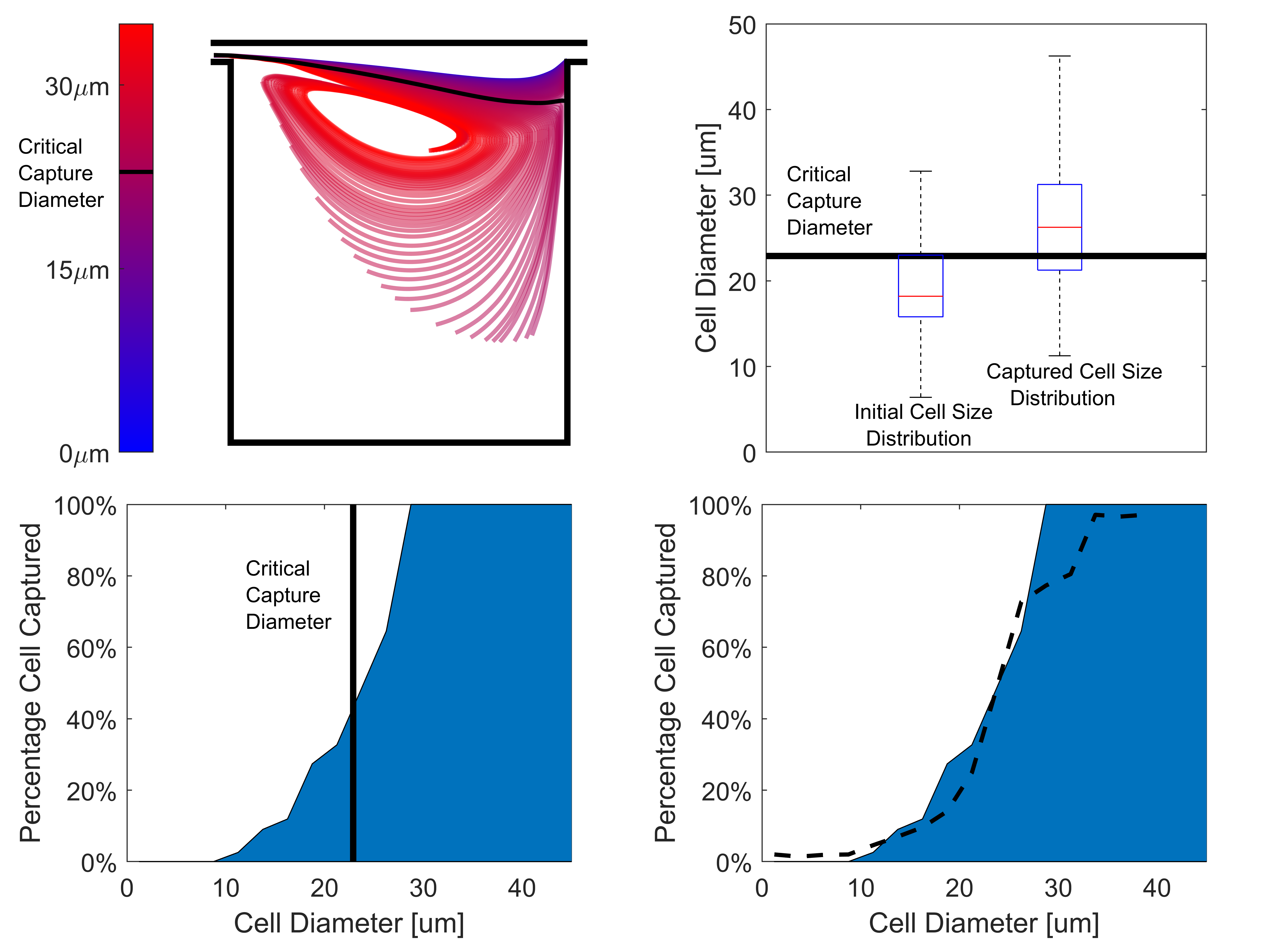}
    \caption{Predicted critical cell diameter for cell sorting matches well with data from micro-centrifuge device in \cite{khojah2017size}. Top left: A continuum of differently sized particle trajectories, smaller particles are in blue and larger particles in red with the critical particle diameter (22.9$\mu $m) is highlighted in black. Top right: Capture data from experiment shows difference in cell size distribution between captured cells and the initial distribution of cells ran through the device. Bottom left: The estimated percentage captured by cell size shows that capture that the beginning of capture is centered around the critical capture diameter. The estimation was necessary because the initial cell size data was given in much coarser bins than the captured cell size data. Bottom right: Estimated percentage captured by cell size is shown along with the percentage captured of trajectories simulation with Eq. \ref{eq:ode} with added Gaussian white noise with $\sigma=12.5\%$ of expected particle velocity.}
    \label{fig:datafig}
\end{figure}

Looking only along the mid plane, we make predictions about the critical Reynolds number for particle capture and compare with experimental data. Kohjoah et al.\cite{khojah2017size} performed an experiment in which a micro-centrifuge was used to separate breast cancer cells (MDA-MB-231 cell line) based on size. Inputting their parameters into our asymptotic method we made a prediction on what the critical cell diameter the device sorts the cells by. Our models predict a critical cell diameter of $22.9\mu $m and this prediction shows good agreement with the data \ref{fig:datafig}.

The experiment consisted of running a blood sample containing 224 cells through the microfluidic chamber 3 times, the experiment was performed twice for a total of N=448 cells run through, with 166 cells captured between the two trials. The device consists of an inlet, a $40 \mu m\times 70\mu$m$ \times 3$cm channel connecting to the $800\mu $m$ \times 70\mu $m$ \times 800\mu $m microcentrifuge chamber, another $40 \mu $m$\times 70 \mu $m$ \times 3 $cm, and an outlet. The distribution of inflowing and captured cells were measured directly from images of the flowing cells, and we use the histograms reported in \cite{khojah2017size} to estimate capture.

We approximated the initial cell size distribution by fitting a log-normal distribution (mean: $20.2\mu $m, standard deviation: $8.2 \mu $m) to available data using the maximum likelihood method. we then approximated the percentage of cells captured in a given size range by dividing the observed number of cells that were captured by the expected number of cells in that size range based on our fitted distribution. For the very largest cells (diameters exceeding $32.75\mu $m) there was insufficient sampling sampling of inflowing particles sizes, leading to capture probabilities that could exceed 1; we set capture probability equal to 1 for these largest cells.

In order to estimate the magnitude of the noise in the system, we fit a modified version of Eq. \ref{eq:ode} with added Gaussian white noise. If we define the RHS of Eq. \ref{eq:ode} as $\mathbf{V}(\alpha, \xp) = \ub (\xp) + \alpha^3\mathcal{M}(\xp) +\frac{\alpha^2}{6}\Delta \ub (\xp)$, our SDE is equal to
\begin{equation}
    \frac{d \xp}{dt} = \mathbf{V}(\alpha, \xp) + ||\mathbf{V}(\alpha, \xp)|| \mathcal{N}(0,\sigma^2)
\end{equation}
We found qualitatively that $\sigma=0.125$ fit the randomness present in the data accurately (fig. \ref{fig:datafig}). Percentage capture of differently sized particles were estimated by computing 1000 trajectories for each level of $\alpha$ and calculating the percentage that were captured.

\subsection{Explanation of Variation}
Our model is deterministic, so all cells above the critical diameter are predicted to be captured and below are not. However, the experimental results reported in \cite{khojah2017size} show that capture probability increase continuously from 0 to 1 as particle radius increases over a narrow range from $10\mu$m to $30\mu$m. The likely cause for this indeterminacy of capture is likely a combination of cells on the top streamline not becoming captured and hydrodynamic interaction between the cells. The micro-centrifuge is intended to be a high throughput device and the cells are focused tightly along the streamlines of the channel feeding the micro-centrifuge chamber. As these cells are near each other, they affect each other's inertial migration due to shear induced dispersion \cite{da1996shear, hood2018pairwise}.

\begin{figure*}
    \centering
    \includegraphics[width=1.0\textwidth]{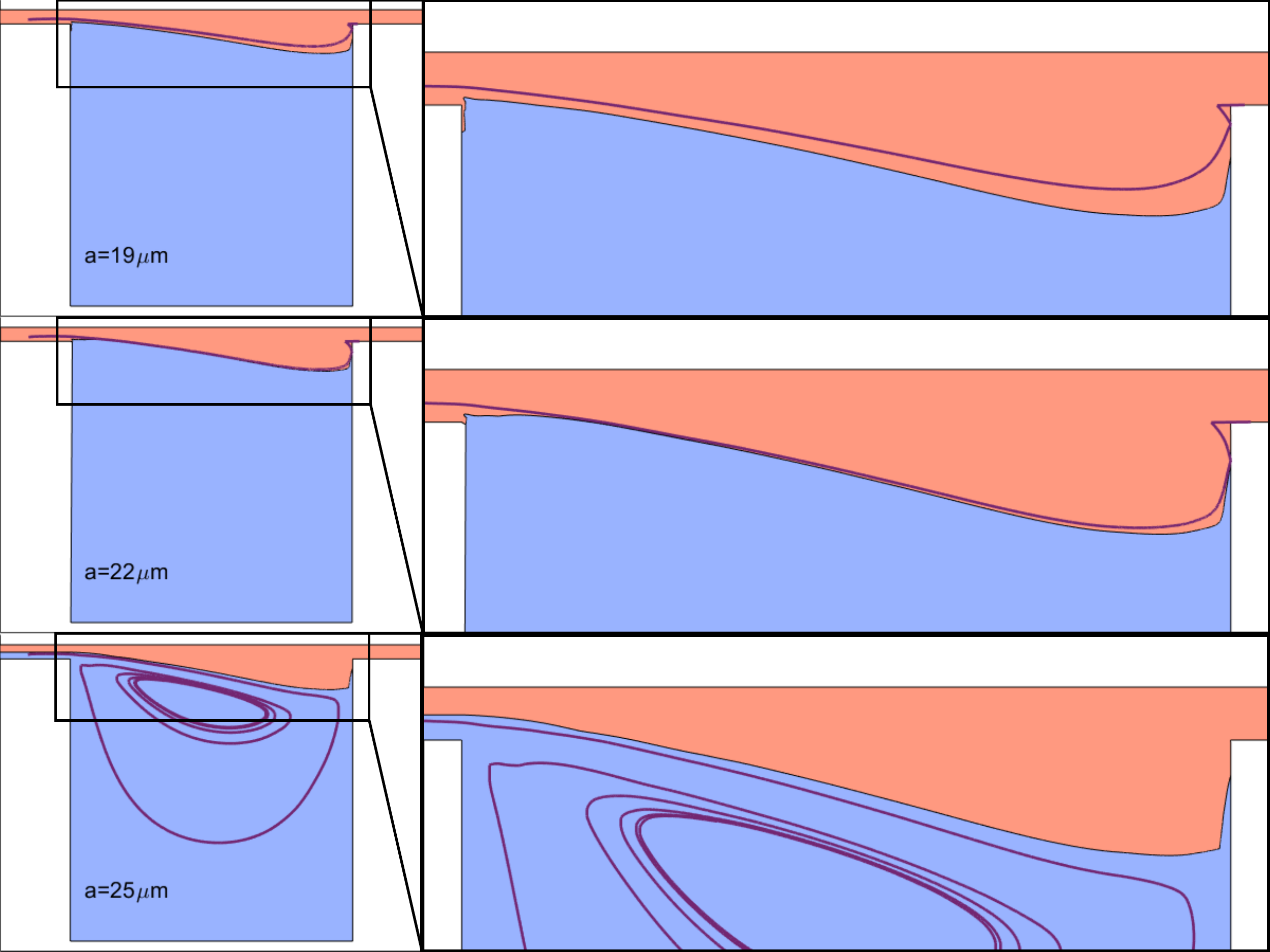}
    \caption{Trajectories of different particles are displayed (purple) over the capture region associated with each sized particle. The red area represents starting points where the particle will not be captured by the vortex and the blue area represents the region where the particle will be captured by the vortex. All trajectories are fairly close to the bifurcating streamline, however trajectories of particles near the critical diameter of $22.9\mu m$ are very close to the separating streamline and whether or not they get captured is influenced by noise.}
    \label{fig:streamline}
\end{figure*}

Inertial focusing is generally the strongest contribution in a particle's cross-stream velocity, so it is not immediately apparent that these unmodeled interactions could be large enough to prevent or ensure capture. However, particle capture trajectories are sensitive to noise. In Fig. \ref{fig:streamline} we visualize the basin of attraction of the microcentrifuge chamber, and we show how close the trajectories of particles approaching the chamber on the lower major focusing streamline are to the separating manifold between the captured region and the un-captured region. In Fig. \ref{fig:streamline} the two colored regions represent invariant sets where a trajectory that starts inside a region will never leave it. 

Each size of particle has different invariant regions because the focusing and curvature term that control particle off-streamline migration in Eq. \ref{eq:ode} change with particle size, however the separating manifold between the two invariant regions is always close to the trajectory of a particle. Particles close to the critical diameter, pass extremely closely to the capturing trajectory for almost the entire length of the chamber, meaning that very small perturbations could permit capture of a particle below the critical capture or deny capture in particles larger than the critical diameter.

\begin{figure}
    \centering
    \includegraphics[width=1\columnwidth]{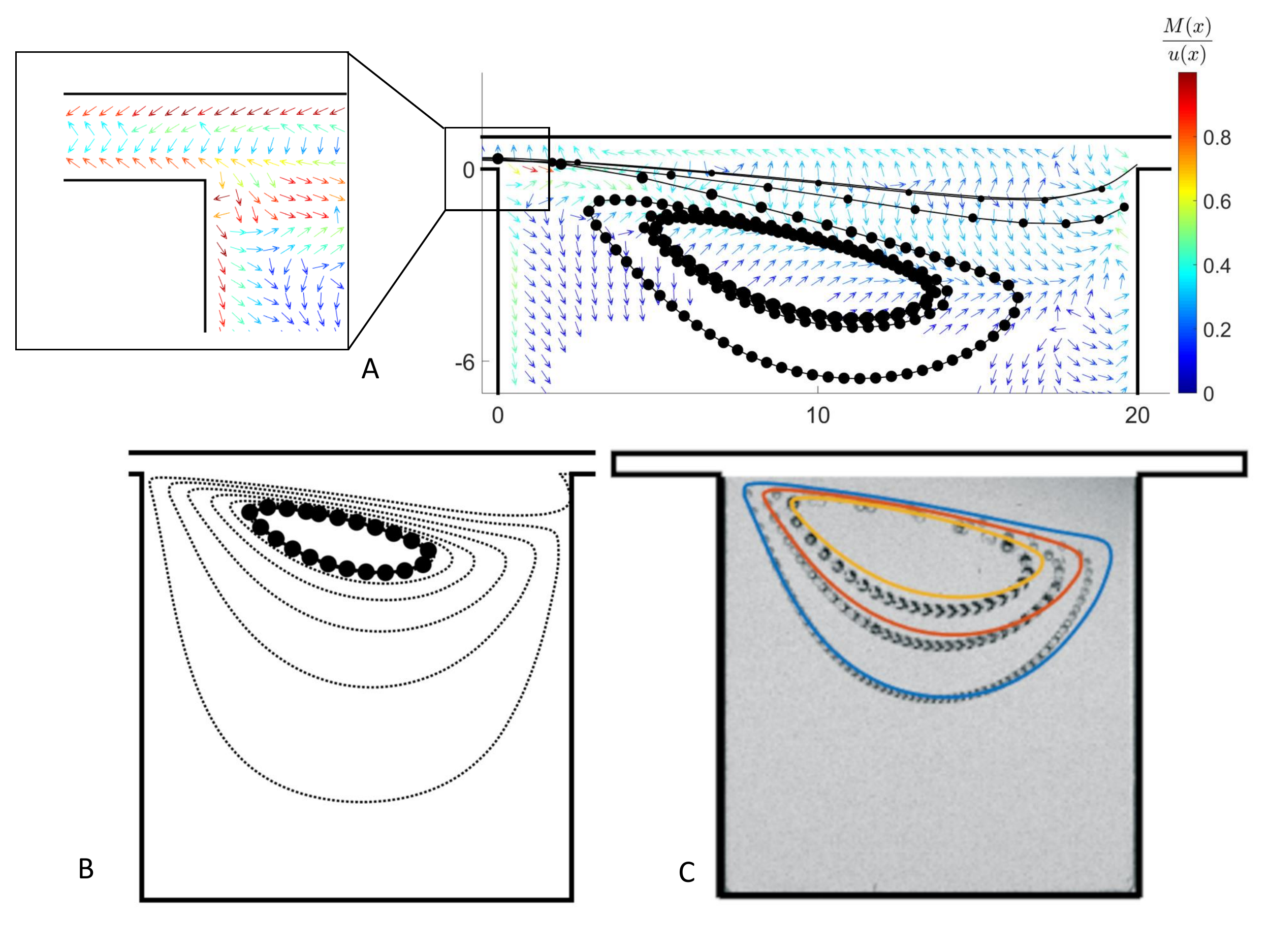}
    \caption{\textbf{A}: A quiver plot of the strength of inertial focusing throughout the channel shown alongside the trajectories of particles of diameters $0\mu m$, $10\mu m$, $20\mu m$, and $30\mu m$.  The strength of migration is related to color, with arrows not shown being $<\frac{1}{100}$ of maximum strength. Strength of inertial migration is measured by $\mathcal{M}(\mathbf{x})/\uu(\mathbf{x})$ which represents the infinitesimally small distance the inertial migration will push the particle at a given point in space. The cutout shows how inertial focusing is strongest in the channel leading to the chamber and near the walls.  \textbf{B}: The particle limit cycle does not trap stream lines, shown with the dotted line, which slowly spiral outwards eventually exiting the microcentrifuge cavity. \textbf{C}: Time lapsed limit cycles from micro-beads of various sizes is overlaid with predictions from our asymptotic simulations. Smaller Particles have larger limit cycles, with the larger particles ($>30\mu m$) having very small limit cycles. While the smaller particles are not predicted to be captured via inertial focusing, their limit cycle is well captured by our asymptotic model.}
    \label{fig:Limitcycle}
\end{figure}

Next we demonstrate why the trajectories of the particles are so close to capturing stream lines using a phase plane diagram showing their inertial migration velocity vectors. this diagram can be scaled to particles of different sizes by scaling the captured velocity vectors by $\alpha^3$ and we superimpose upon the vector field plot the trajectories of three different particle sizes. We see that the inertial migration is strongest in the channel leading up to the micro-centrifuge and fairly weak near the vortex itself. This is because inertial migration depends upon shear, curvature, and walls \cite{di2009inertial}. Near the vortex the shear is low so the inertial migration is low. The trajectories of particles near the critical particle diameter all experience very similar inertial migration, the larger particles are captured because they experience slightly more. We are unaware of any experimental observations of particle trajectories during capture, or more generally, for particles crossing the microcentrifuge and this experimental gap is likely due to the high speed of particles within the channels feeding each cavity. By contrast, the limiting trajectories of captured particles within the channels have been extensively reported on. It is known that captured particles are drawn into closed orbits (limit cycles) within the micro centrifuge; our simulations replicate these dynamics (Fig. \ref{fig:Limitcycle}A) and agree quantitatively with measured limit cycle paths for particles of different sizes at $Re=125$ (Fig. \ref{fig:Limitcycle}) \cite{mach2011automated, zhou2013enhanced, haddadi2017inertial}. In particular we find that larger particles converge to smaller orbits. Orbit shape is well predicted, including a common center of gyration, a long straight trajectory that is tilted and closely conforming to the separating streamline that divides captured and non-captured particles (Fig. \ref{fig:streamline}). To fit experimental data we had to compute particle sizes from the figures of \cite{khojah2017size}. We chose diameters $15.3\mu m$, $16.2\mu m$, and $20.5\mu m$ because they fit the sizes listed in out parts of \cite{khojah2017size}. The relative sizes we measured in figure were 8.1, 8.6, and 12.2 pixels respectively. There are some discrepancies, though it is not possible to determine whether they are due to imperfect matching of the Reynolds number or particle sizes, unmodeled effects such as the device flexing slightly when under pressure or whether they are due to approximations made in our model. 

The two smaller particles are not predicted to be captured by our simulation, but nevertheless they are predicted to have a stable limit cycle if they enter the cavity. This feature complicates the calculation of capture probabilities; small particles do not migrate across the separating manifold in Fig \ref{fig:streamline}, but if perturbations, such as due to multi-particle effects, induce them to cross it they will be stably entrained in the vortex eddy. The smallest particle that is predicted to have a stable limit cycle in our simulation had a diameter of $10.9 \mu $m which lines up well with data where the smallest cell captured had diameter between $10-12.5\mu$m. the complicating presence of these smaller particles within the microcentrifuge chamber emphasizes that the vortex is unrelated to the capture of particles, the vortex only controls the retention of particles.

\subsection{How to Change Critical Particle Diameter}
While the calculations results above are a significant analysis of the particles, they cannot be extended to other channels as the background flow depends on the Reynolds number. However, if we change the channel size and flow speed to keep our Reynolds number constant, we can reuse all of our calculations but with a different parameter set, allowing us to select for the critical particle diameter that our microfluidic device selects for:

\begin{eqnarray}
Re &=& \frac{UL}{\nu} = \frac{\frac{U}{k}kL}{\nu} \nonumber \\
\alpha &=& a/kL \nonumber \\
\alpha^*_k &=& \alpha^*/k \nonumber
\end{eqnarray}
For example, if you need to change the device so that instead of separating particles above and below $22.9\mu m$ to selecting particles above and below $11.45\mu m$, all you need to do is half the size of your device and double your speed.
\\
\\
There is a practical limitation however, the PDMS polymer that the microfluidic devices are etched from do not come at an arbitrary thickness. However, there have been previous studies that if you scale the fluid speed and channel/cavity height accordingly the change in thickness does not effect the fluid flow \cite{https://doi.org/10.48550/arxiv.2101.07242}. We speculate, although leave the testing to future work that this change is also unlikely to effect the inertial migration significantly, that as the optimal Reynolds number for separation drives particles to the symmetry plane, the furthest place from the side walls.

\section{Perspectives}

By replacing particle's with singularities, we were able to accurately model inertial migration without solving the nonlinear Navier-Stokes equations or explicitly meshing the 3D and time-varying fluid domain between particles and channel walls. This method allowed us carefully analyze all possible particle trajectories within a microcentrifuge device. Our analysis shows two clear takeaways for microfluidic devices designed to separate particles by size:
\begin{enumerate}
     \item The preformance of the microcentrifuge is deeply dependent on Reynolds number. Lower Reynolds numbers advect particles towards the symmetry plane of the device and allow for more stable separation of particles by size. Higher Reynolds numbers push particles away from the symmetry plane of the device and introduce more complicated 3D effects, which are both difficult to anatomize and have been found empirically to cause devices to be have wider, less consistent capture thresholds. 
    \item Error in sorting is due to lack of separation between particle trajectories and the capture separatrix. Particles within $3\mu$m of the critical capture diameter are straddling the capture separatrix for 75\% the length of the microcentrifuge, making their capture susceptible to particle-particle interaction and other forms of random noise. Inertial migration velocity scale as $O(\alpha^3)$, the high order size scaling leads to sufficient separation for sorting particle accurately by size, but is not enough to have low error rates by itself.
\end{enumerate}
Experimental studies that have optimized for capture accuracy have independently agreed with this design principles. \cite{khojah2017size,haddadi2017inertial,wang2013vortex} have all found optimal Reynolds numbers between 100-150 and \cite{wang2013vortex} were able to achieve 90\% capture accuracy at 90\% purity by optimizing over Reynolds number. We identify the effect of this optimization as finding maximum separation between focusing position in the channel and separating manifold.

Our singularity method models only the first order inertial migration, further expansions in terms of particle size and Reynolds number are possible, and would add either higher order forcing terms within our Oseen equation, or would require that we impose additional matching conditions to model the particle. Such an extension could be useful for predicting differential focusing of different particles within the same channel. Unpacking the role of particle interactions and extending the calculation to the largest particles on which separation is performed. The simulations here took approximately 10 minutes per solve of inertial migration, however computes could be sped up significantly by using discontinuity decompositions from \cite{christensen2022fast} to . The method can be extended to particles of other shapes, singularity modeling of the particle needs only the stresslet strength associated with the Stokes solution around the particle. This stresslet strength is already known for ellipsoids in shear flow\cite{chwang1975hydromechanics} and for other particles can be found by solving for the motion of the particle in Stokes flow, for which there are many approximate or numerical methods \cite{happel2012low,kim2013microhydrodynamics}.

Changes to channel geometry can also lead to significant increases in separation accuracy. Pai\'e et al. \cite{paie2017effect} found a 20\% increase in accuracy of capture by creating a separate reservoir for captured particles and optimizing over 6 parameter combinations. Numerically driven parameter optimization is still extremely computationally expensive, here we have done our best to analyze indepth the basic microcentrifuge geometry and describe the physical principles that that fuel its strengths and weaknesses. Our results provide a platform on which accelerated studies may be used to further optimize the accuracy, controllability, or throughput of future microcentrifuges.



\nocite{*}

\bibliography{bibMain}

\providecommand{\noopsort}[1]{}\providecommand{\singleletter}[1]{#1}%
\begin{thebibliography}{32}%
\makeatletter
\providecommand \@ifxundefined [1]{%
 \@ifx{#1\undefined}
}%
\providecommand \@ifnum [1]{%
 \ifnum #1\expandafter \@firstoftwo
 \else \expandafter \@secondoftwo
 \fi
}%
\providecommand \@ifx [1]{%
 \ifx #1\expandafter \@firstoftwo
 \else \expandafter \@secondoftwo
 \fi
}%
\providecommand \natexlab [1]{#1}%
\providecommand \enquote  [1]{``#1''}%
\providecommand \bibnamefont  [1]{#1}%
\providecommand \bibfnamefont [1]{#1}%
\providecommand \citenamefont [1]{#1}%
\providecommand \href@noop [0]{\@secondoftwo}%
\providecommand \href [0]{\begingroup \@sanitize@url \@href}%
\providecommand \@href[1]{\@@startlink{#1}\@@href}%
\providecommand \@@href[1]{\endgroup#1\@@endlink}%
\providecommand \@sanitize@url [0]{\catcode `\\12\catcode `\$12\catcode
  `\&12\catcode `\#12\catcode `\^12\catcode `\_12\catcode `\%12\relax}%
\providecommand \@@startlink[1]{}%
\providecommand \@@endlink[0]{}%
\providecommand \url  [0]{\begingroup\@sanitize@url \@url }%
\providecommand \@url [1]{\endgroup\@href {#1}{\urlprefix }}%
\providecommand \urlprefix  [0]{URL }%
\providecommand \Eprint [0]{\href }%
\providecommand \doibase [0]{https://doi.org/}%
\providecommand \selectlanguage [0]{\@gobble}%
\providecommand \bibinfo  [0]{\@secondoftwo}%
\providecommand \bibfield  [0]{\@secondoftwo}%
\providecommand \translation [1]{[#1]}%
\providecommand \BibitemOpen [0]{}%
\providecommand \bibitemStop [0]{}%
\providecommand \bibitemNoStop [0]{.\EOS\space}%
\providecommand \EOS [0]{\spacefactor3000\relax}%
\providecommand \BibitemShut  [1]{\csname bibitem#1\endcsname}%
\let\auto@bib@innerbib\@empty
\bibitem [{\citenamefont {Khojah}\ \emph {et~al.}(2017)\citenamefont {Khojah},
  \citenamefont {Stoutamore},\ and\ \citenamefont {Di~Carlo}}]{khojah2017size}%
  \BibitemOpen
  \bibfield  {author} {\bibinfo {author} {\bibfnamefont {R.}~\bibnamefont
  {Khojah}}, \bibinfo {author} {\bibfnamefont {R.}~\bibnamefont {Stoutamore}},\
  and\ \bibinfo {author} {\bibfnamefont {D.}~\bibnamefont {Di~Carlo}},\
  }\bibfield  {title} {\bibinfo {title} {Size-tunable microvortex capture of
  rare cells},\ }\href@noop {} {\bibfield  {journal} {\bibinfo  {journal} {Lab
  on a Chip}\ }\textbf {\bibinfo {volume} {17}},\ \bibinfo {pages} {2542}
  (\bibinfo {year} {2017})}\BibitemShut {NoStop}%
\bibitem [{\citenamefont {Haddadi}\ and\ \citenamefont
  {Di~Carlo}(2017)}]{haddadi2017inertial}%
  \BibitemOpen
  \bibfield  {author} {\bibinfo {author} {\bibfnamefont {H.}~\bibnamefont
  {Haddadi}}\ and\ \bibinfo {author} {\bibfnamefont {D.}~\bibnamefont
  {Di~Carlo}},\ }\bibfield  {title} {\bibinfo {title} {Inertial flow of a
  dilute suspension over cavities in a microchannel},\ }\href@noop {}
  {\bibfield  {journal} {\bibinfo  {journal} {Journal of Fluid Mechanics}\
  }\textbf {\bibinfo {volume} {811}},\ \bibinfo {pages} {436} (\bibinfo {year}
  {2017})}\BibitemShut {NoStop}%
\bibitem [{\citenamefont {Che}\ \emph {et~al.}(2016)\citenamefont {Che},
  \citenamefont {Yu}, \citenamefont {Dhar}, \citenamefont {Renier},
  \citenamefont {Matsumoto}, \citenamefont {Heirich}, \citenamefont {Garon},
  \citenamefont {Goldman}, \citenamefont {Rao}, \citenamefont {Sledge} \emph
  {et~al.}}]{che2016classification}%
  \BibitemOpen
  \bibfield  {author} {\bibinfo {author} {\bibfnamefont {J.}~\bibnamefont
  {Che}}, \bibinfo {author} {\bibfnamefont {V.}~\bibnamefont {Yu}}, \bibinfo
  {author} {\bibfnamefont {M.}~\bibnamefont {Dhar}}, \bibinfo {author}
  {\bibfnamefont {C.}~\bibnamefont {Renier}}, \bibinfo {author} {\bibfnamefont
  {M.}~\bibnamefont {Matsumoto}}, \bibinfo {author} {\bibfnamefont
  {K.}~\bibnamefont {Heirich}}, \bibinfo {author} {\bibfnamefont {E.~B.}\
  \bibnamefont {Garon}}, \bibinfo {author} {\bibfnamefont {J.}~\bibnamefont
  {Goldman}}, \bibinfo {author} {\bibfnamefont {J.}~\bibnamefont {Rao}},
  \bibinfo {author} {\bibfnamefont {G.~W.}\ \bibnamefont {Sledge}}, \emph
  {et~al.},\ }\bibfield  {title} {\bibinfo {title} {Classification of large
  circulating tumor cells isolated with ultra-high throughput microfluidic
  vortex technology},\ }\href@noop {} {\bibfield  {journal} {\bibinfo
  {journal} {Oncotarget}\ }\textbf {\bibinfo {volume} {7}},\ \bibinfo {pages}
  {12748} (\bibinfo {year} {2016})}\BibitemShut {NoStop}%
\bibitem [{\citenamefont {Dhar}\ \emph {et~al.}(2018)\citenamefont {Dhar},
  \citenamefont {Wong}, \citenamefont {Che}, \citenamefont {Matsumoto},
  \citenamefont {Grogan}, \citenamefont {Elashoff}, \citenamefont {Garon},
  \citenamefont {Goldman}, \citenamefont {Christen}, \citenamefont {Di~Carlo}
  \emph {et~al.}}]{dhar2018evaluation}%
  \BibitemOpen
  \bibfield  {author} {\bibinfo {author} {\bibfnamefont {M.}~\bibnamefont
  {Dhar}}, \bibinfo {author} {\bibfnamefont {J.}~\bibnamefont {Wong}}, \bibinfo
  {author} {\bibfnamefont {J.}~\bibnamefont {Che}}, \bibinfo {author}
  {\bibfnamefont {M.}~\bibnamefont {Matsumoto}}, \bibinfo {author}
  {\bibfnamefont {T.}~\bibnamefont {Grogan}}, \bibinfo {author} {\bibfnamefont
  {D.}~\bibnamefont {Elashoff}}, \bibinfo {author} {\bibfnamefont {E.~B.}\
  \bibnamefont {Garon}}, \bibinfo {author} {\bibfnamefont {J.~W.}\ \bibnamefont
  {Goldman}}, \bibinfo {author} {\bibfnamefont {E.~S.}\ \bibnamefont
  {Christen}}, \bibinfo {author} {\bibfnamefont {D.}~\bibnamefont {Di~Carlo}},
  \emph {et~al.},\ }\bibfield  {title} {\bibinfo {title} {Evaluation of pd-l1
  expression on vortex-isolated circulating tumor cells in metastatic lung
  cancer},\ }\href@noop {} {\bibfield  {journal} {\bibinfo  {journal}
  {Scientific reports}\ }\textbf {\bibinfo {volume} {8}},\ \bibinfo {pages} {1}
  (\bibinfo {year} {2018})}\BibitemShut {NoStop}%
\bibitem [{\citenamefont {Chen}\ \emph {et~al.}(2012)\citenamefont {Chen},
  \citenamefont {Li},\ and\ \citenamefont {Sun}}]{chen2012microfluidic}%
  \BibitemOpen
  \bibfield  {author} {\bibinfo {author} {\bibfnamefont {J.}~\bibnamefont
  {Chen}}, \bibinfo {author} {\bibfnamefont {J.}~\bibnamefont {Li}},\ and\
  \bibinfo {author} {\bibfnamefont {Y.}~\bibnamefont {Sun}},\ }\bibfield
  {title} {\bibinfo {title} {Microfluidic approaches for cancer cell detection,
  characterization, and separation},\ }\href@noop {} {\bibfield  {journal}
  {\bibinfo  {journal} {Lab on a Chip}\ }\textbf {\bibinfo {volume} {12}},\
  \bibinfo {pages} {1753} (\bibinfo {year} {2012})}\BibitemShut {NoStop}%
\bibitem [{\citenamefont {Schonberg}\ and\ \citenamefont
  {Hinch}(1989)}]{schonberg1989inertial}%
  \BibitemOpen
  \bibfield  {author} {\bibinfo {author} {\bibfnamefont {J.~A.}\ \bibnamefont
  {Schonberg}}\ and\ \bibinfo {author} {\bibfnamefont {E.}~\bibnamefont
  {Hinch}},\ }\bibfield  {title} {\bibinfo {title} {Inertial migration of a
  sphere in poiseuille flow},\ }\href@noop {} {\bibfield  {journal} {\bibinfo
  {journal} {Journal of Fluid Mechanics}\ }\textbf {\bibinfo {volume} {203}},\
  \bibinfo {pages} {517} (\bibinfo {year} {1989})}\BibitemShut {NoStop}%
\bibitem [{\citenamefont {Hood}\ \emph {et~al.}(2015)\citenamefont {Hood},
  \citenamefont {Lee},\ and\ \citenamefont {Roper}}]{hood_lee_roper_2015}%
  \BibitemOpen
  \bibfield  {author} {\bibinfo {author} {\bibfnamefont {K.}~\bibnamefont
  {Hood}}, \bibinfo {author} {\bibfnamefont {S.}~\bibnamefont {Lee}},\ and\
  \bibinfo {author} {\bibfnamefont {M.}~\bibnamefont {Roper}},\ }\bibfield
  {title} {\bibinfo {title} {Inertial migration of a rigid sphere in
  three-dimensional poiseuille flow},\ }\href
  {https://doi.org/10.1017/jfm.2014.739} {\bibfield  {journal} {\bibinfo
  {journal} {Journal of Fluid Mechanics}\ }\textbf {\bibinfo {volume} {765}},\
  \bibinfo {pages} {452–479} (\bibinfo {year} {2015})}\BibitemShut {NoStop}%
\bibitem [{\citenamefont {Sierou}\ and\ \citenamefont
  {Brady}(2001)}]{sierou2001accelerated}%
  \BibitemOpen
  \bibfield  {author} {\bibinfo {author} {\bibfnamefont {A.}~\bibnamefont
  {Sierou}}\ and\ \bibinfo {author} {\bibfnamefont {J.~F.}\ \bibnamefont
  {Brady}},\ }\bibfield  {title} {\bibinfo {title} {Accelerated stokesian
  dynamics simulations},\ }\href@noop {} {\bibfield  {journal} {\bibinfo
  {journal} {Journal of fluid mechanics}\ }\textbf {\bibinfo {volume} {448}},\
  \bibinfo {pages} {115} (\bibinfo {year} {2001})}\BibitemShut {NoStop}%
\bibitem [{\citenamefont {Liron}\ and\ \citenamefont
  {Barta}(1992)}]{liron1992motion}%
  \BibitemOpen
  \bibfield  {author} {\bibinfo {author} {\bibfnamefont {N.}~\bibnamefont
  {Liron}}\ and\ \bibinfo {author} {\bibfnamefont {E.}~\bibnamefont {Barta}},\
  }\bibfield  {title} {\bibinfo {title} {Motion of a rigid particle in stokes
  flow: a new second-kind boundary-integral equation formulation},\ }\href@noop
  {} {\bibfield  {journal} {\bibinfo  {journal} {Journal of Fluid Mechanics}\
  }\textbf {\bibinfo {volume} {238}},\ \bibinfo {pages} {579} (\bibinfo {year}
  {1992})}\BibitemShut {NoStop}%
\bibitem [{\citenamefont {Kazerooni}\ \emph {et~al.}(2017)\citenamefont
  {Kazerooni}, \citenamefont {Fornari}, \citenamefont {Hussong},\ and\
  \citenamefont {Brandt}}]{kazerooni2017inertial}%
  \BibitemOpen
  \bibfield  {author} {\bibinfo {author} {\bibfnamefont {H.~T.}\ \bibnamefont
  {Kazerooni}}, \bibinfo {author} {\bibfnamefont {W.}~\bibnamefont {Fornari}},
  \bibinfo {author} {\bibfnamefont {J.}~\bibnamefont {Hussong}},\ and\ \bibinfo
  {author} {\bibfnamefont {L.}~\bibnamefont {Brandt}},\ }\bibfield  {title}
  {\bibinfo {title} {Inertial migration in dilute and semidilute suspensions of
  rigid particles in laminar square duct flow},\ }\href@noop {} {\bibfield
  {journal} {\bibinfo  {journal} {Physical Review Fluids}\ }\textbf {\bibinfo
  {volume} {2}},\ \bibinfo {pages} {084301} (\bibinfo {year}
  {2017})}\BibitemShut {NoStop}%
\bibitem [{\citenamefont {Nakagawa}\ \emph {et~al.}(2015)\citenamefont
  {Nakagawa}, \citenamefont {Yabu}, \citenamefont {Otomo}, \citenamefont
  {Kase}, \citenamefont {Makino}, \citenamefont {Itano},\ and\ \citenamefont
  {Sugihara-Seki}}]{nakagawa2015inertial}%
  \BibitemOpen
  \bibfield  {author} {\bibinfo {author} {\bibfnamefont {N.}~\bibnamefont
  {Nakagawa}}, \bibinfo {author} {\bibfnamefont {T.}~\bibnamefont {Yabu}},
  \bibinfo {author} {\bibfnamefont {R.}~\bibnamefont {Otomo}}, \bibinfo
  {author} {\bibfnamefont {A.}~\bibnamefont {Kase}}, \bibinfo {author}
  {\bibfnamefont {M.}~\bibnamefont {Makino}}, \bibinfo {author} {\bibfnamefont
  {T.}~\bibnamefont {Itano}},\ and\ \bibinfo {author} {\bibfnamefont
  {M.}~\bibnamefont {Sugihara-Seki}},\ }\bibfield  {title} {\bibinfo {title}
  {Inertial migration of a spherical particle in laminar square channel flows
  from low to high reynolds numbers},\ }\href@noop {} {\bibfield  {journal}
  {\bibinfo  {journal} {Journal of Fluid Mechanics}\ }\textbf {\bibinfo
  {volume} {779}},\ \bibinfo {pages} {776} (\bibinfo {year}
  {2015})}\BibitemShut {NoStop}%
\bibitem [{\citenamefont {Xu}\ \emph {et~al.}(2020)\citenamefont {Xu},
  \citenamefont {Gao}, \citenamefont {Lofquist}, \citenamefont {Fernando},
  \citenamefont {Hsu}, \citenamefont {Sundar},\ and\ \citenamefont
  {Ganapathysubramanian}}]{xu2020octree}%
  \BibitemOpen
  \bibfield  {author} {\bibinfo {author} {\bibfnamefont {S.}~\bibnamefont
  {Xu}}, \bibinfo {author} {\bibfnamefont {B.}~\bibnamefont {Gao}}, \bibinfo
  {author} {\bibfnamefont {A.}~\bibnamefont {Lofquist}}, \bibinfo {author}
  {\bibfnamefont {M.}~\bibnamefont {Fernando}}, \bibinfo {author}
  {\bibfnamefont {M.-C.}\ \bibnamefont {Hsu}}, \bibinfo {author} {\bibfnamefont
  {H.}~\bibnamefont {Sundar}},\ and\ \bibinfo {author} {\bibfnamefont
  {B.}~\bibnamefont {Ganapathysubramanian}},\ }\bibfield  {title} {\bibinfo
  {title} {An octree-based immersogeometric approach for modeling inertial
  migration of particles in channels},\ }\href@noop {} {\bibfield  {journal}
  {\bibinfo  {journal} {Computers \& Fluids}\ ,\ \bibinfo {pages} {104764}}
  (\bibinfo {year} {2020})}\BibitemShut {NoStop}%
\bibitem [{\citenamefont {Bazaz}\ \emph {et~al.}(2020)\citenamefont {Bazaz},
  \citenamefont {Mashhadian}, \citenamefont {Ehsani}, \citenamefont {Saha},
  \citenamefont {Kr{\"u}ger},\ and\ \citenamefont
  {Warkiani}}]{bazaz2020computational}%
  \BibitemOpen
  \bibfield  {author} {\bibinfo {author} {\bibfnamefont {S.~R.}\ \bibnamefont
  {Bazaz}}, \bibinfo {author} {\bibfnamefont {A.}~\bibnamefont {Mashhadian}},
  \bibinfo {author} {\bibfnamefont {A.}~\bibnamefont {Ehsani}}, \bibinfo
  {author} {\bibfnamefont {S.~C.}\ \bibnamefont {Saha}}, \bibinfo {author}
  {\bibfnamefont {T.}~\bibnamefont {Kr{\"u}ger}},\ and\ \bibinfo {author}
  {\bibfnamefont {M.~E.}\ \bibnamefont {Warkiani}},\ }\bibfield  {title}
  {\bibinfo {title} {Computational inertial microfluidics: a review},\
  }\href@noop {} {\bibfield  {journal} {\bibinfo  {journal} {Lab on a Chip}\
  }\textbf {\bibinfo {volume} {20}},\ \bibinfo {pages} {1023} (\bibinfo {year}
  {2020})}\BibitemShut {NoStop}%
\bibitem [{\citenamefont {Sun}\ \emph {et~al.}(2016)\citenamefont {Sun},
  \citenamefont {Wang}, \citenamefont {Dong},\ and\ \citenamefont
  {Sun}}]{sun2016three}%
  \BibitemOpen
  \bibfield  {author} {\bibinfo {author} {\bibfnamefont {D.-K.}\ \bibnamefont
  {Sun}}, \bibinfo {author} {\bibfnamefont {Y.}~\bibnamefont {Wang}}, \bibinfo
  {author} {\bibfnamefont {A.-P.}\ \bibnamefont {Dong}},\ and\ \bibinfo
  {author} {\bibfnamefont {B.-D.}\ \bibnamefont {Sun}},\ }\bibfield  {title}
  {\bibinfo {title} {A three-dimensional quantitative study on the hydrodynamic
  focusing of particles with the immersed boundary--lattice boltzmann method},\
  }\href@noop {} {\bibfield  {journal} {\bibinfo  {journal} {International
  Journal of Heat and Mass Transfer}\ }\textbf {\bibinfo {volume} {94}},\
  \bibinfo {pages} {306} (\bibinfo {year} {2016})}\BibitemShut {NoStop}%
\bibitem [{\citenamefont {Chun}\ and\ \citenamefont
  {Ladd}(2006)}]{chun2006inertial}%
  \BibitemOpen
  \bibfield  {author} {\bibinfo {author} {\bibfnamefont {B.}~\bibnamefont
  {Chun}}\ and\ \bibinfo {author} {\bibfnamefont {A.}~\bibnamefont {Ladd}},\
  }\bibfield  {title} {\bibinfo {title} {Inertial migration of neutrally
  buoyant particles in a square duct: An investigation of multiple equilibrium
  positions},\ }\href@noop {} {\bibfield  {journal} {\bibinfo  {journal}
  {Physics of Fluids}\ }\textbf {\bibinfo {volume} {18}},\ \bibinfo {pages}
  {031704} (\bibinfo {year} {2006})}\BibitemShut {NoStop}%
\bibitem [{\citenamefont {Loisel}\ \emph {et~al.}(2013)\citenamefont {Loisel},
  \citenamefont {Abbas}, \citenamefont {Masbernat},\ and\ \citenamefont
  {Climent}}]{loisel2013effect}%
  \BibitemOpen
  \bibfield  {author} {\bibinfo {author} {\bibfnamefont {V.}~\bibnamefont
  {Loisel}}, \bibinfo {author} {\bibfnamefont {M.}~\bibnamefont {Abbas}},
  \bibinfo {author} {\bibfnamefont {O.}~\bibnamefont {Masbernat}},\ and\
  \bibinfo {author} {\bibfnamefont {E.}~\bibnamefont {Climent}},\ }\bibfield
  {title} {\bibinfo {title} {The effect of neutrally buoyant finite-size
  particles on channel flows in the laminar-turbulent transition regime},\
  }\href@noop {} {\bibfield  {journal} {\bibinfo  {journal} {Physics of
  Fluids}\ }\textbf {\bibinfo {volume} {25}},\ \bibinfo {pages} {123304}
  (\bibinfo {year} {2013})}\BibitemShut {NoStop}%
\bibitem [{\citenamefont {Abbas}\ \emph {et~al.}(2014)\citenamefont {Abbas},
  \citenamefont {Magaud}, \citenamefont {Gao},\ and\ \citenamefont
  {Geoffroy}}]{abbas2014migration}%
  \BibitemOpen
  \bibfield  {author} {\bibinfo {author} {\bibfnamefont {M.}~\bibnamefont
  {Abbas}}, \bibinfo {author} {\bibfnamefont {P.}~\bibnamefont {Magaud}},
  \bibinfo {author} {\bibfnamefont {Y.}~\bibnamefont {Gao}},\ and\ \bibinfo
  {author} {\bibfnamefont {S.}~\bibnamefont {Geoffroy}},\ }\bibfield  {title}
  {\bibinfo {title} {Migration of finite sized particles in a laminar square
  channel flow from low to high reynolds numbers},\ }\href@noop {} {\bibfield
  {journal} {\bibinfo  {journal} {Physics of Fluids}\ }\textbf {\bibinfo
  {volume} {26}},\ \bibinfo {pages} {123301} (\bibinfo {year}
  {2014})}\BibitemShut {NoStop}%
\bibitem [{\citenamefont {Kim}\ and\ \citenamefont
  {Karrila}(2013)}]{kim2013microhydrodynamics}%
  \BibitemOpen
  \bibfield  {author} {\bibinfo {author} {\bibfnamefont {S.}~\bibnamefont
  {Kim}}\ and\ \bibinfo {author} {\bibfnamefont {S.~J.}\ \bibnamefont
  {Karrila}},\ }\href@noop {} {\emph {\bibinfo {title} {Microhydrodynamics:
  principles and selected applications}}}\ (\bibinfo  {publisher} {Courier
  Corporation},\ \bibinfo {year} {2013})\BibitemShut {NoStop}%
\bibitem [{\citenamefont {Multiphysics}(1998)}]{multiphysics1998introduction}%
  \BibitemOpen
  \bibfield  {author} {\bibinfo {author} {\bibfnamefont {C.}~\bibnamefont
  {Multiphysics}},\ }\bibfield  {title} {\bibinfo {title} {Introduction to
  comsol multiphysics{\textregistered}},\ }\href@noop {} {\bibfield  {journal}
  {\bibinfo  {journal} {COMSOL Multiphysics, Burlington, MA, accessed Feb}\
  }\textbf {\bibinfo {volume} {9}},\ \bibinfo {pages} {2018} (\bibinfo {year}
  {1998})}\BibitemShut {NoStop}%
\bibitem [{\citenamefont {Di~Carlo}\ \emph {et~al.}(2007)\citenamefont
  {Di~Carlo}, \citenamefont {Irimia}, \citenamefont {Tompkins},\ and\
  \citenamefont {Toner}}]{di2007continuous}%
  \BibitemOpen
  \bibfield  {author} {\bibinfo {author} {\bibfnamefont {D.}~\bibnamefont
  {Di~Carlo}}, \bibinfo {author} {\bibfnamefont {D.}~\bibnamefont {Irimia}},
  \bibinfo {author} {\bibfnamefont {R.~G.}\ \bibnamefont {Tompkins}},\ and\
  \bibinfo {author} {\bibfnamefont {M.}~\bibnamefont {Toner}},\ }\bibfield
  {title} {\bibinfo {title} {Continuous inertial focusing, ordering, and
  separation of particles in microchannels},\ }\href@noop {} {\bibfield
  {journal} {\bibinfo  {journal} {Proceedings of the National Academy of
  Sciences}\ }\textbf {\bibinfo {volume} {104}},\ \bibinfo {pages} {18892}
  (\bibinfo {year} {2007})}\BibitemShut {NoStop}%
\bibitem [{\citenamefont {Christensen}\ \emph {et~al.}(2022)\citenamefont
  {Christensen}, \citenamefont {Chu}, \citenamefont {Anderson},\ and\
  \citenamefont {Roper}}]{christensen2022fast}%
  \BibitemOpen
  \bibfield  {author} {\bibinfo {author} {\bibfnamefont {S.}~\bibnamefont
  {Christensen}}, \bibinfo {author} {\bibfnamefont {R.}~\bibnamefont {Chu}},
  \bibinfo {author} {\bibfnamefont {C.}~\bibnamefont {Anderson}},\ and\
  \bibinfo {author} {\bibfnamefont {M.}~\bibnamefont {Roper}},\ }\bibfield
  {title} {\bibinfo {title} {Fast asymptotic-numerical method for coarse mesh
  particle simulation in channels of arbitrary cross section},\ }\href@noop {}
  {\bibfield  {journal} {\bibinfo  {journal} {Journal of Computational
  Physics}\ }\textbf {\bibinfo {volume} {471}},\ \bibinfo {pages} {111622}
  (\bibinfo {year} {2022})}\BibitemShut {NoStop}%
\bibitem [{\citenamefont {Khojah}\ \emph {et~al.}(2021)\citenamefont {Khojah},
  \citenamefont {Lo}, \citenamefont {Tang},\ and\ \citenamefont
  {Di~Carlo}}]{https://doi.org/10.48550/arxiv.2101.07242}%
  \BibitemOpen
  \bibfield  {author} {\bibinfo {author} {\bibfnamefont {R.}~\bibnamefont
  {Khojah}}, \bibinfo {author} {\bibfnamefont {D.}~\bibnamefont {Lo}}, \bibinfo
  {author} {\bibfnamefont {F.}~\bibnamefont {Tang}},\ and\ \bibinfo {author}
  {\bibfnamefont {D.}~\bibnamefont {Di~Carlo}},\ }\href
  {https://doi.org/10.48550/ARXIV.2101.07242} {\bibinfo {title} {The evolution
  of flow and mass transport in 3d confined cavities}} (\bibinfo {year}
  {2021})\BibitemShut {NoStop}%
\bibitem [{\citenamefont {Da~Cunha}\ and\ \citenamefont
  {Hinch}(1996)}]{da1996shear}%
  \BibitemOpen
  \bibfield  {author} {\bibinfo {author} {\bibfnamefont {F.}~\bibnamefont
  {Da~Cunha}}\ and\ \bibinfo {author} {\bibfnamefont {E.}~\bibnamefont
  {Hinch}},\ }\bibfield  {title} {\bibinfo {title} {Shear-induced dispersion in
  a dilute suspension of rough spheres},\ }\href@noop {} {\bibfield  {journal}
  {\bibinfo  {journal} {Journal of fluid mechanics}\ }\textbf {\bibinfo
  {volume} {309}},\ \bibinfo {pages} {211} (\bibinfo {year}
  {1996})}\BibitemShut {NoStop}%
\bibitem [{\citenamefont {Hood}\ and\ \citenamefont
  {Roper}(2018)}]{hood2018pairwise}%
  \BibitemOpen
  \bibfield  {author} {\bibinfo {author} {\bibfnamefont {K.}~\bibnamefont
  {Hood}}\ and\ \bibinfo {author} {\bibfnamefont {M.}~\bibnamefont {Roper}},\
  }\bibfield  {title} {\bibinfo {title} {Pairwise interactions in inertially
  driven one-dimensional microfluidic crystals},\ }\href@noop {} {\bibfield
  {journal} {\bibinfo  {journal} {Physical Review Fluids}\ }\textbf {\bibinfo
  {volume} {3}},\ \bibinfo {pages} {094201} (\bibinfo {year}
  {2018})}\BibitemShut {NoStop}%
\bibitem [{\citenamefont {Di~Carlo}(2009)}]{di2009inertial}%
  \BibitemOpen
  \bibfield  {author} {\bibinfo {author} {\bibfnamefont {D.}~\bibnamefont
  {Di~Carlo}},\ }\bibfield  {title} {\bibinfo {title} {Inertial
  microfluidics},\ }\href@noop {} {\bibfield  {journal} {\bibinfo  {journal}
  {Lab on a Chip}\ }\textbf {\bibinfo {volume} {9}},\ \bibinfo {pages} {3038}
  (\bibinfo {year} {2009})}\BibitemShut {NoStop}%
\bibitem [{\citenamefont {Mach}\ \emph {et~al.}(2011)\citenamefont {Mach},
  \citenamefont {Kim}, \citenamefont {Arshi}, \citenamefont {Hur},\ and\
  \citenamefont {Di~Carlo}}]{mach2011automated}%
  \BibitemOpen
  \bibfield  {author} {\bibinfo {author} {\bibfnamefont {A.~J.}\ \bibnamefont
  {Mach}}, \bibinfo {author} {\bibfnamefont {J.~H.}\ \bibnamefont {Kim}},
  \bibinfo {author} {\bibfnamefont {A.}~\bibnamefont {Arshi}}, \bibinfo
  {author} {\bibfnamefont {S.~C.}\ \bibnamefont {Hur}},\ and\ \bibinfo {author}
  {\bibfnamefont {D.}~\bibnamefont {Di~Carlo}},\ }\bibfield  {title} {\bibinfo
  {title} {Automated cellular sample preparation using a
  centrifuge-on-a-chip},\ }\href@noop {} {\bibfield  {journal} {\bibinfo
  {journal} {Lab on a Chip}\ }\textbf {\bibinfo {volume} {11}},\ \bibinfo
  {pages} {2827} (\bibinfo {year} {2011})}\BibitemShut {NoStop}%
\bibitem [{\citenamefont {Zhou}\ \emph {et~al.}(2013)\citenamefont {Zhou},
  \citenamefont {Kasper},\ and\ \citenamefont {Papautsky}}]{zhou2013enhanced}%
  \BibitemOpen
  \bibfield  {author} {\bibinfo {author} {\bibfnamefont {J.}~\bibnamefont
  {Zhou}}, \bibinfo {author} {\bibfnamefont {S.}~\bibnamefont {Kasper}},\ and\
  \bibinfo {author} {\bibfnamefont {I.}~\bibnamefont {Papautsky}},\ }\bibfield
  {title} {\bibinfo {title} {Enhanced size-dependent trapping of particles
  using microvortices},\ }\href@noop {} {\bibfield  {journal} {\bibinfo
  {journal} {Microfluidics and nanofluidics}\ }\textbf {\bibinfo {volume}
  {15}},\ \bibinfo {pages} {611} (\bibinfo {year} {2013})}\BibitemShut
  {NoStop}%
\bibitem [{\citenamefont {Wang}\ \emph {et~al.}(2013)\citenamefont {Wang},
  \citenamefont {Zhou},\ and\ \citenamefont {Papautsky}}]{wang2013vortex}%
  \BibitemOpen
  \bibfield  {author} {\bibinfo {author} {\bibfnamefont {X.}~\bibnamefont
  {Wang}}, \bibinfo {author} {\bibfnamefont {J.}~\bibnamefont {Zhou}},\ and\
  \bibinfo {author} {\bibfnamefont {I.}~\bibnamefont {Papautsky}},\ }\bibfield
  {title} {\bibinfo {title} {Vortex-aided inertial microfluidic device for
  continuous particle separation with high size-selectivity, efficiency, and
  purity},\ }\href@noop {} {\bibfield  {journal} {\bibinfo  {journal}
  {Biomicrofluidics}\ }\textbf {\bibinfo {volume} {7}},\ \bibinfo {pages}
  {044119} (\bibinfo {year} {2013})}\BibitemShut {NoStop}%
\bibitem [{\citenamefont {Chwang}\ and\ \citenamefont
  {Wu}(1975)}]{chwang1975hydromechanics}%
  \BibitemOpen
  \bibfield  {author} {\bibinfo {author} {\bibfnamefont {A.~T.}\ \bibnamefont
  {Chwang}}\ and\ \bibinfo {author} {\bibfnamefont {T.}~\bibnamefont {Wu}},\
  }\bibfield  {title} {\bibinfo {title} {Hydromechanics of low-reynolds-number
  flow. part 2. singularity method for stokes flows},\ }\href@noop {}
  {\bibfield  {journal} {\bibinfo  {journal} {Journal of Fluid mechanics}\
  }\textbf {\bibinfo {volume} {67}},\ \bibinfo {pages} {787} (\bibinfo {year}
  {1975})}\BibitemShut {NoStop}%
\bibitem [{\citenamefont {Happel}\ and\ \citenamefont
  {Brenner}(2012)}]{happel2012low}%
  \BibitemOpen
  \bibfield  {author} {\bibinfo {author} {\bibfnamefont {J.}~\bibnamefont
  {Happel}}\ and\ \bibinfo {author} {\bibfnamefont {H.}~\bibnamefont
  {Brenner}},\ }\href@noop {} {\emph {\bibinfo {title} {Low Reynolds number
  hydrodynamics: with special applications to particulate media}}},\
  Vol.~\bibinfo {volume} {1}\ (\bibinfo  {publisher} {Springer Science \&
  Business Media},\ \bibinfo {year} {2012})\BibitemShut {NoStop}%
\bibitem [{\citenamefont {Pai{\`e}}\ \emph {et~al.}(2017)\citenamefont
  {Pai{\`e}}, \citenamefont {Che},\ and\ \citenamefont
  {Di~Carlo}}]{paie2017effect}%
  \BibitemOpen
  \bibfield  {author} {\bibinfo {author} {\bibfnamefont {P.}~\bibnamefont
  {Pai{\`e}}}, \bibinfo {author} {\bibfnamefont {J.}~\bibnamefont {Che}},\ and\
  \bibinfo {author} {\bibfnamefont {D.}~\bibnamefont {Di~Carlo}},\ }\bibfield
  {title} {\bibinfo {title} {Effect of reservoir geometry on vortex trapping of
  cancer cells},\ }\href@noop {} {\bibfield  {journal} {\bibinfo  {journal}
  {Microfluidics and Nanofluidics}\ }\textbf {\bibinfo {volume} {21}},\
  \bibinfo {pages} {1} (\bibinfo {year} {2017})}\BibitemShut {NoStop}%
\bibitem [{\citenamefont {Haddadi}\ \emph {et~al.}(2018)\citenamefont
  {Haddadi}, \citenamefont {Naghsh-Nilchi},\ and\ \citenamefont
  {Di~Carlo}}]{haddadi2018separation}%
  \BibitemOpen
  \bibfield  {author} {\bibinfo {author} {\bibfnamefont {H.}~\bibnamefont
  {Haddadi}}, \bibinfo {author} {\bibfnamefont {H.}~\bibnamefont
  {Naghsh-Nilchi}},\ and\ \bibinfo {author} {\bibfnamefont {D.}~\bibnamefont
  {Di~Carlo}},\ }\bibfield  {title} {\bibinfo {title} {Separation of cancer
  cells using vortical microfluidic flows},\ }\href@noop {} {\bibfield
  {journal} {\bibinfo  {journal} {Biomicrofluidics}\ }\textbf {\bibinfo
  {volume} {12}},\ \bibinfo {pages} {014112} (\bibinfo {year}
  {2018})}\BibitemShut {NoStop}%
\end{thebibliography}%

\end{document}